\documentstyle[prd,aps,epsfig]{revtex} 
\begin{document}
\preprint{UMDPP #00-038}

\twocolumn[\hsize\textwidth\columnwidth\hsize\csname
@twocolumnfalse\endcsname
\draft

\title{Unraveling critical dynamics:  The formation and evolution of topological textures}
\author{G. J. Stephens}

\address{Department of Physics, University of Maryland\\
College Park, Maryland 20742-4111\\
gstephen@physics.umd.edu} 

\maketitle
\widetext
\begin{abstract}
We study the formation of topological textures in a nonequilibrium phase transition
of an overdamped classical O(3) model in $2+1$ dimensions.  The phase transition 
is triggered through an external, time-dependent effective mass, parameterized by quench timescale $\tau$. 
When measured near the end of the transition ($\langle \vec{\Phi}^2 \rangle=0.9$) the average 
texture separation $L_{sep}$ and the average texture width $L_w$ scale as
 $L_{sep} \sim \tau^{0.39 \pm 0.02}$ and $L_w \sim \tau^{0.46 \pm 0.04}$, 
significantly larger than the single power-law $\xi_{freeze} \sim \tau^{0.25}$
predicted from the Kibble-Zurek mechanism.  We show that Kibble-Zurek scaling is recovered at
very early times but that by the end of the transition $L_{sep}(\tau)$ and $L_w(\tau)$ result instead
from a competition between the length scale determined at freeze-out and the ordering dynamics of a 
textured system.  
We offer a simple proposal for the dynamics of these length scales:  
$L_{sep}(t)=\xi_1(\tau)+L_1(t-t_{freeze})$ and $L_{w}(t)=\xi_2 (\tau)+L_2(t-t_{freeze})$ 
where $\xi_1 \sim \tau^\alpha$ and $\xi_2 \sim \tau^\beta$ are determined by the freeze-out
mechanism.  $L_1(t)=(\xi_1)^{\frac{1}{3}}t^{\frac{1}{3}}$ and $L_2(t)=t^{\frac{1}{2}}$ are dynamical length scales 
previously known from phase ordering dynamics, and $t_{freeze}$ is the freeze-out time.  
We find that $L_{sep}(t)$ and $L_w(t)$ fit closely to the length scales observed at the
end of the transition and yield $\alpha=0.24 \pm 0.02$ and $\beta=0.22 \pm 0.07$, in good
agreement with the Kibble-Zurek mechanism.  In the context of phase ordering these 
results suggest that the multiple length scales characteristic of the late-time ordering of a 
textured system derive from the critical dynamics of a {\it single} nonequilibrium correlation length. 
In the context of defect formation these results imply that significant evolution of
the defect network can occur before the end of the phase transition.  Therefore a quantitative understanding
of the defect network at the end of the phase transition generally requires an understanding of both critical
dynamics {\it and} the interactions among topological defects. 
\end{abstract}
\pacs{PACS number(s): 05.70.Ln, 11.27.+d ,64.60Ht, 12.39.D
\hfill UMDPP\#00-038}

\narrowtext
\vskip2pc]

\section{Introduction}

We live in a world of broken symmetry.  From the crystalline structure of ice, 
no longer invariant under the full rotational and translational symmetry of liquid water, 
to the vacuum state of the standard model, the universe is filled with the
patterns of symmetry breaking.  Understanding the dynamical processes by 
which symmetry-breaking phase transitions occur is therefore of substantial physical importance.  Topological
defects, forged in nonlinear interactions as the system seeks its true vacuum, offer
a window into the complex, nonequilibrium aspects of the phase transition.    
In systems with a topologically nontrivial vacuum manifold, fields in
different spatial regions fall into different ground states and symmetry breaking may be
accompanied by the emergence of a well-defined network of topological defects at the end of the 
transition \cite{Kib}.  Simple topological defects include domain walls, strings, monopoles and 
topological textures and are themselves the subject of much research in both 
cosmology and the laboratory \cite{VilShel} \cite{Mer}.  Systems
containing topological textures are of particular interest as they provide an arena in which to 
explore myriad dynamical issues ranging from nonequilibrium phase transitions to the coarsening dynamics of 
quantum field theory. 

\subsection{Topological defects in cosmology and the laboratory}
  
Very little is known about the exact nature of the early universe before nucleosynthesis.  However, many
particle phenomenologies predict a unification of different forces at high energies, $T_{GUT}\sim 10^{16} GeV$.
If such unification exits then a rich pattern of symmetry-breaking phase transitions developed as the  
universe expanded and cooled.  Phase transitions also occurred at the lower
electroweak and QCD energy scales, although important physical details such as the exact order of
the transitions remain unknown.

In the course of symmetry breaking, causality constrains the correlation length to grow no larger than the 
size of the horizon. Regions of space smaller than a horizon volume will rest in 
different vacua. For systems in which the vacuum 
manifold $M$ has nontrivial homotopy groups $\Pi_n(M)$, topological defects may exist where different vacua meet.  
Topological defects formed in the early universe imprint themselves upon
the fluctuations of the cosmic microwave background radiation (CMBR) and have been suggested 
as the generators of large scale structure formation (for a review see \cite{VilShel}).  Much work has been done 
to compare measurements of the multipole moments of the CMBR anisotropies and numerical 
simulations of defect models.  While the apparent lack of agreement between theory and observations is not
immeadiately promising, the noisy nature of current measurements and the errors involved in doing large-scale 
numerical simulations of cosmological defect evolution make it difficult to rule out defect models of
structure formation with any certainty (see for example \cite{DurKunzMel}).  

The study of topological defects in the early universe has been fruitful
not only in attempting to explain precise cosmological data, but also in influencing the direction of 
cosmological research. One of the early motivations for the paradigm of inflation was the predicted overabundance of
magnetic monopoles \cite{Guth}.  If monopoles are formed in a phase transition before or during 
an inflationary phase, their number 
density is safely diluted by the exponentially expanding scale factor.  
However, it has also recently been observed that in the very energetic process of reheating, as coherent
oscillations of the inflaton decay into other fields, a nonthermal spectrum of topological
defects may be produced, thus again raising the monopole problem \cite{KofLin}.  The details of the 
reheating process and the possible formation of a defect distribution at the end of an inflationary 
phase are currently an area of active study \cite{Reheat}.

Nonequilibrium phase transitions and the formation of topological defects are observed
in many laboratory systems and the study of topological defects has helped forge links
between the study of high energy processes in the early universe and condensed matter systems \cite{Zur}.  
In liquid crystals,
the formation of a network of topological string defects has been observed 
in the rapid cooling of the system through its critical point, demonstrating not only the viability of the 
Kibble mechanism in producing topological defects but also the scaling of the defect network \cite{LiqCry}.  
Other condensed matter systems such as superfluid $^3$He, have complex symmetry-breaking patterns and offer 
many potential analogs to early universe processes \cite{Vol}.  

Topological textures form in $d$ spatial dimensions when the vacuum manifold $M$ is such that $\Pi_d(M)$ is nontrivial.
Unlike other topological defects, textured fields never leave the vacuum manifold but instead are ``knots'' of
gradient energy, given relative stability by their topological winding around the vacuum.  Like cosmic strings, 
the energy density in cosmological textures scales with the expansion of the universe so that textures 
formed in early universe phase transitions do not come to dominate the matter density. For this reason,  
global textures have been studied extensively as candidates for generators of large-scale 
structure formation \cite{Tur1,Tur2}.  In condensed matter, topological textures are studied in a 
wide variety of (mostly two-dimensional) systems, from superconductors \cite{supertex} and quantum hall 
ferromagnets \cite{qhtex} to superfluid $^3He-A$ \cite{hetex}. 

\subsection{Nonequilibrium phase transitions and topological textures}
The network of topological defects produced during a nonequilibrium phase transition contains dynamical 
information about the evolution of the phase transition.  A rapid phase transition can
create a topological defect density far above equilibrium values.   The defect
density decreases as the system equilibrates after the transition.  However the topological stability of defects 
allows any defect overdensity to last for a substantial period of time.   Current understanding of the 
dynamics of topological defect formation is summarized by the Kibble-Zurek mechanism which describes
the emergence of a {\it single} nonequilibrium length scale $\xi_{freeze}$ characteristic
of the initial defect density \cite{Zur}. The freeze-out scale $\xi_{freeze}$ is
derived by combining phenomenological ideas of near-equilibrium critical dynamics and equilibrium critical scaling
and predicts a power-law dependence of the number of defects with the quench rate of the phase transition.
Both theoretical \cite{DefTheory} \cite{ZurLag,Jac,ZurYat,AntBetZur,SteCalHuRam} and 
experimental \cite{DefExp,KibExp} effort has been focused on exploring the connection  between the 
length scale of the initial defect distribution and the quench rate of the phase transition.  
These researches study the formation of topological defects that contain false vacuum in their cores:  
kinks in one dimension and vortices in 2 and 3 dimensions.  Although (as explained below) the dynamics of a 
textured system can be qualitatively different, the details of the formation of topological textures in a 
nonequilibrium phase transition are mostly unexplored (see however \cite{Jac}). 

The dynamics of textured systems provides an ideal study for nonequilibrium phase transitions.
Unlike other topological defects textures do not have a fixed size and the evolution of a textured system 
is complicated both by the poorly understood interactions between textures and by the changing 
scale of the texture itself.  The richness of texture interactions involves multiple dynamical
length scales.  In $2+1$ dimensions it is observed that the phase-ordering of a system containing 
topological textures is characterized at late times by at least 
{\it three} length scales: the (average) texture size, 
texture-texture separation and texture-antitexture separation \cite{ZapZak,Rut}.  In distinction, systems of 
topological defects such as domain walls, vortices or monopoles are effectively single-scale and are 
described at late-times only by the average defect separation.  In a nonequilibrium phase transition
the multiple independent length scales of textured systems are determined by the phase transition dynamics.  
However, the arguments of the Kibble-Zurek mechanism are ambiguous when applied to textured systems.
Which, if any, of the multiple length scales observed in a texture distribution should we equate 
with the ``frozen'' correlation length?

\subsection{Organization}

Section II reviews the properties of topological textures in $2+1$ dimensions and 
the scaling violations observed in textured systems when quenched from a disordered phase.  In section III,
we present the Kibble-Zurek mechanism and derive the single nonequilibrium freeze-out scale $\xi_{freeze}$ 
characteristic of the
defect network at the time of formation.  Section IV contains the model of a nonequilibrium phase transition, 
the numerical techniques used in the evolution and a description of the evolution of the phase transition given by
the numerical solution.  We also define the length scales used to characterize the texture distribution. 
Section V contains the main results of this work: an explanation of the quench rate 
dependence of the average texture separation and the average texture width observed near the end of 
the phase transition.  We show that the Kibble-Zurek mechanism is recovered at early times but that by the end of the
phase transition, formation dynamics and phase ordering dynamics are intrinsically linked.  
In Section VI these results are used to argue that textured systems carry an imprint of the freeze-out 
correlation length to late times and suggest this offers the possibility of novel, 
late-time measurements of dynamical critical phenomena.  Concluding remarks are made in Section VII.

\section{Topological Textures in two spatial dimensions}

The properties of systems containing topological textures differ in many ways from those containing
other topological defects.  To illustrate these differences, consider an O(3) invariant model with classical action,

\begin{equation}
\label{eq-action}
S=\int d^3x [\partial_{\mu}\vec{\Phi}\cdot\partial^{\mu}\vec{\Phi}+\frac{\lambda}{4}(|\vec{\Phi}|^2-m^2)^2],
\end{equation}

\noindent and resulting equations of motion

\begin{equation}
\label{eq-fieldeq1}
\ddot{\Phi}_i-\bigtriangledown^2\Phi_i+\lambda\Phi_i(\vec{\Phi}^2-m^2)=0.
\end{equation}

\noindent For the classical dynamics that is of interest here the value of the quartic self-coupling $\lambda$ and
the mass parameter $m^2$ may be scaled to unity with an appropriate scaling of the field and spacetime units,

\begin{equation}
\Phi_i \rightarrow \Phi_i m^2, \quad t \rightarrow t \sqrt{\lambda} m, \quad \vec{x} \rightarrow \vec{x} \sqrt{\lambda} m.
\end{equation}

\noindent The vacuum manifold is characterized by $\vec{\Phi}\cdot \vec{\Phi}=1$, topologically a
2-sphere. When the field is constrained to the vacuum the action is that of a nonlinear sigma model (NLSM). 
For asymptotically uniform fields, a vacuum field configuration is a map 
\begin{equation}
\vec{\Phi}: S^2_{real \thinspace  space} \rightarrow S^2_{field \thinspace space}, 
\end{equation}
\noindent divided into distinct topological sectors by winding number $n$ \cite{BelPol}.  In two spatial dimensions these
windings are topological textures, topologically stable and static solutions with finite energy. For example, 
the following field configuration, a texture of negative unit winding, wraps around the vacuum 2-sphere exactly once, 
\begin{equation}
\label{eq-texture}
\Phi_1=\frac{4ar\sin(\phi)}{r^3+4a^2},\quad \Phi_2=\frac{4ar\cos(\phi)}{r^2+4a^2},\quad  \Phi_3=\frac{r^2-4a^2}{r^2+4a^2}.
\end{equation}
\noindent At the origin, $\vec{\Phi}$ points in the $-\Phi_3$ direction while for $r\rightarrow \infty$ it points in 
the opposite direction, $\Phi_3$.  In between, at $r=2a$, the field points radially outward like a hedgehog.  
The parameter $a$ characterizes the size of the texture and is independent of the parameters of the 
action Eq. (\ref{eq-action}).  The NLSM also possesses an exactly conserved 
topological charge $Q$, which is expressed as the integral over a  
topological charge density $\rho$,

\begin{eqnarray}
\rho=\frac{1}{4\pi}\vec{\Phi}\cdot (\partial_x\vec{\Phi}\wedge \partial_y\vec{\Phi})\\
Q=\int d^2x \rho(x,y)
\end{eqnarray}

\noindent For the single winding texture, Eq. (\ref{eq-texture}), the topological charge density has a particularly simple form
\begin{equation}
\rho=-\frac{1}{\pi}\frac{4a^2}{(r^2+4a^2)^2},
\end{equation}

\noindent with total topological charge $Q=-1$. The energy $E=\frac{1}{2} 
\int d^2x (\vec{\partial}\Phi_i \cdot \vec{\partial}\Phi_i)$
of the single winding configuration Eq. (\ref{eq-texture}) is simply $E=8\pi$, independent of the size $a$ of the
texture.  In accordance with Derricks theorem \cite{Der}, the energy of a texture configuration in higher dimensions 
scales as a positive power of its size, demonstrating their instability to collapse.  In lower dimensions textures are 
unstable to expansion.  The stability of a texture configuration in any dimension can be altered by adding higher derivative
terms to action Eq. (\ref{eq-action}), as was done in the Skyrme model of nucleons \cite{Skyrme}.  In contrast to 
other topological defects such as strings or monopoles, the texture core size $a$ is 
variable and independent of the fundamental parameters of the theory.

Although the energy of a single isolated topological texture is independent of its size, systems with multiple
textures are not static.  Textured systems {\it order} under the constraint of conserved topological charge Q.
However, the details of texture interactions are not well understood.  
It is observed that textures and antitextures can annihilate with each other and that more isolated textures can
decay by unwinding \cite{ZapZak,Rut}.  The details of the unwinding process depend on the model under
consideration.  For the action Eq.(\ref{eq-action}) textures can decay by pulling the field off the
vacuum manifold and unwinding.  This is analogous to isolated textures in higher dimensions which are unstable 
to shrinking and unwind when their gradient energy is large enough to move them off the vacuum manifold.  
In a NLSM where the fields must remain on the vacuum manifold, textures can unwind through
higher-derivative terms present in the lattice discretization.    

\subsection{Topological textures and the violation of dynamical scaling}

When a disordered system is quenched into the ordered phase (e.g. by the removal of thermal fluctuations)
the approach to a new equilibrium and associated long-range order is through a sequence of nonequilibrium states.
A simple example is the Ising ferromagnet in 3 spatial dimension.  Above the Curie temperature, spins are
randomly oriented and there is no net magnetization.  A snapshot immeadiately following the quench of this system 
to zero temperature shows a system containing many domains in which the spins are coherently aligned.  
The net magnetization remains zero because the domains are randomly oriented with respect to each other. 
However, the true $T=0$ ground state of this system consists of only one
domain and no domain walls.  The system approaches this equilibrium through the dynamical process of 
coarsening during which small domains shrink and large domains grow.  The process of coarsening has been studied 
extensively in both condensed matter systems and in the early universe (for reviews see \cite{Bray} and \cite{HinKib}).  
The {\it scaling hypothesis} arose from these studies and states that, at late times, in a quench from a 
disordered state, the domain distribution is statistically characterized by a {\it single} dynamical scale $L(t)$ 
which grows in time.  For simple dissipative systems, the approach to ($T=0$) equilibrium  is 
governed by model-A time-dependent Landau-Ginzburg (TDGL) dynamics for a nonconserved order parameter 
\begin{equation}
\label{eq-tdgl}
\frac {\partial \Phi(x,t)} {\partial t}=-\Gamma \frac{\delta F[\Phi]} {\delta \Phi}.
\end{equation}
It has been shown both theoretically and experimentally that the average domain size $L(t)$
grows with time as

\begin{equation}
\label{eq-scaling}
 L(t)=(t\Gamma)^\frac{1}{2},
\end {equation}

\noindent which is also the power-law predicted by a dimensional analysis of Eq. (\ref{eq-tdgl}).
Not all systems coarsen in such a simple way and there is currently no general theoretical framework 
in which to explain why some systems scale while others do not.  In particular, systems containing 
topological textures in one and two spatial dimensions are strong exceptions to the 
scaling hypothesis  \cite{ZapZak,Rut,RutBray}.  In a $2+1$ dimensional O(3) model with dissipative
dynamics and an instantaneous quench from the disordered phase, late-time coarsening of 
topological textures is described by at least {\it three} different length scales \cite{ZapZak,Rut}. 
These three length scales are the average defect-defect separation $L_{sep}$, the average defect width $L_w$, and the
average texture-antitexture separation $L_{tat}$.  The scaling hypothesis is violated since
at late times these three length scales are observed to grow with different powers of time

\begin{eqnarray}
L_{sep}(t)=\xi_0^\frac{1}{3}t^\frac {1}{3} \\
L_w(t)=\xi_0^\frac{2}{3} t^\frac {1}{6} \\
L_{tat}(t)= t^\frac {1}{2}
\end{eqnarray}

\noindent where $\xi_0$ is the correlation length in the disordered phase.  It is remarkable that length
scales describing the late-time ordering of textures retain a dependence on the
initial correlation length.  In distinction, systems that obey the 
scaling hypothesis erase any memory of their initial state and at late times
retain only a {\it dynamical} scale (Eq. (\ref{eq-scaling}) for the case of the ferromagnet). 
The scaling of systems that contain topological defects such as vortices implies that 
the vortex distribution at late times is independent of the dynamical details of the phase transition.  
For textured systems it is {\it not} possible to have such a clean separation between the dynamics of 
the phase transition and the late-time texture distribution.
In the context of a nonequilibrium phase transition $\xi_0$ is determined {\it not} from
initial conditions but from the Kibble-Zurek mechanism, thus connecting early-time critical dynamics and 
late-time coarsening.  In Section VII we discus how this connection may provide late-time 
probes of nonequilibrium critical dynamics.

\section{The Kibble-Zurek Mechanism of Defect Formation}

Before analyzing the dynamics of topological texture formation it is useful to review the basic Kibble-Zurek
mechanism. The first estimate of the initial defect density in a cosmological context was made
by Kibble \cite{Kib}.  The basic ingredients of the 
Kibble mechanism are causality and the Ginzburg temperature, $T_G$.  The 
Ginzburg temperature is defined as the temperature at which thermal fluctuations contain just enough energy for 
correlated regions of the field to overcome the potential energy barrier
between inequivalent vacua,
 
\begin {equation}
k_bT_G \sim \xi(T_G)^3 \Delta F(T_G),
\end {equation}

\noindent  where $\Delta F$ is the difference in free energy density between the true and false vacua and
 $\xi$ is the equilibrium correlation length.
In the Kibble mechanism, the length scale characterizing the initial defect network is set by the equilibrium 
correlation length of the field, evaluated at $T_G$.
In a recent series of experiments \cite{KibExp}, the Kibble mechanism was tested in the laboratory.  The results, while
confirming the production of defects in a symmetry-breaking phase transition, 
indicate that 
$\xi(T_G)$ does {\it not} set the characteristic length scale of the initial defect distribution.  These
experiments were suggested by Zurek, who criticized Kibble's
use of equilibrium arguments and the Ginzburg temperature, $T_G$ \cite {Zur}.
  
Combining equilibrium and non equilibrium ingredients, Zurek offers a
``freeze-out'' proposal to estimate the initial density of defects. 
In the Zurek proposal, above the critical temperature, the field starts off in thermal equilibrium with a heat bath.  As the 
temperature of the bath is lowered adiabatically, the field remains in local thermal equilibrium with the heat bath. Near 
the phase transition,  
the equilibrium correlation 
length and the equilibrium relaxation time of the field grow without bound as  

\begin {eqnarray}
\xi = \xi_0 {\mid \epsilon \mid}^{-\nu}, \\
\tau_r = \tau_0 {\mid \epsilon \mid}^{-\mu}, 
\end {eqnarray}

\noindent where $\epsilon$ characterizes the proximity to the critical temperature,
\begin {equation}
\label{eq-epsilon}
\epsilon = \frac {T_c-T} {T_c}, 
\end {equation}

\noindent and $\mu$ and $\nu$ are critical exponents appropriate for the theory under consideration.  
The quench is assumed to occur linearly in time

\begin {equation}
\epsilon=\frac {t} {\tau}
\end {equation}

\noindent so that for $t<0$, the temperature of the heat bath is above the critical temperature
and the critical temperature is reached at $t=0$. 
The divergence of the equilibrium relaxation time as the heat bath approaches the critical temperature
is known as critical slowing down.
Critical slowing down results from the finite speed of propagation of perturbations of the order 
parameter.  As the correlation length diverges, small 
perturbations of the order parameter ({\it e.g.,} lowering of the temperature) take longer to propagate over correlated 
regions and therefore it takes longer to reach equilibrium.  As the critical temperature is approached from above there 
comes a time 
$\mid t* \mid $ during the quench when the time remaining before the 
transition equals the equilibrium relaxation time

\begin {equation}
 \mid t* \mid=\tau_r(t*).
\end {equation}

\noindent Beyond this point the correlation length can 
no longer adjust fast enough to follow the changing temperature of the bath.  At time {\it t*} the dynamics of the
correlation length ``freezes".   The correlation 
length remains frozen until a time {\it t*} after the critical temperature is reached.  
In Zurek's proposal the 
correlation length at the ``freeze-out'' time {\it t*} sets the characteristic length scale
for the initial defect network.  
Solving for the value of the correlation length at {\it t*}, the frozen correlation length, 
and therefore the initial defect density, scale with the quench rate as

\begin {equation}
\xi (t*) \sim {\tau}^{\frac {\nu} {1+\mu}}.
\end {equation}

\noindent The power-law scaling of the topological defect density with the quench rate has been verified 
using 1-dimensional, 2-dimensional and 3-dimensional simulations of 
phenomenological time-dependent Landau-Ginzburg equations \cite{ZurLag,Jac,ZurYat,AntBetZur} and in
nonequilibrium quantum field theory \cite{SteCalHuRam}.  However, experimental efforts to test the Kibble-Zurek prediction in 
condensed matter systems are inconclusive, lacking reliable error estimates and a mechanism to vary 
the quench rate \cite{DefExp}.  In particular, quenches in $^4He$ and high-temperature
superconductors do not seem to produce defect densities at the level expected from the freeze-out mechanism.

\section {Critical Dynamics}

In order to study the {\it dynamical} formation of topological texture the equations of motion Eq. (\ref{eq-fieldeq1}) are
modified with an externally controlled time-dependent effective mass,

\begin{equation}
m^2_{eff}(t)=\tanh(\frac{t_c-t}{\tau}),
\end{equation}

\noindent where $\tau$ is the quench rate and $t_c$ defines the critical point where 
spinodal instabilities, characteristic of a second order phase transition, begin to grow.  Both
in the early universe and in the laboratory, texture formation occurs in an environment containing many
other degrees of freedom.  The effect of the environment is {\it assumed} to be that
of a stochastic (white) noise driving term $\xi(x,t)$ and simple ohmic dissipation, obeying a fluctuation-dissipation
relation.  The field equations of motion are    

\begin{equation}
\label{eq-fieldeq2}
\ddot{\Phi}_i-\bigtriangledown^2\Phi_i-\eta\dot{\Phi}_i+\Phi_i(\vec{\Phi}^2-m^2_{eff})=\xi(\vec{x},t),
\end{equation}

\noindent where the fluctuation-dissipation relation is
\begin{equation} 
\label{eq-fd}
\langle \xi(\vec{x},t) \xi(\vec{x}',t') \rangle = 2T\eta\delta^2(\vec{x}-\vec{x}')\delta(t-t').
\end{equation}

\noindent The use of a time-dependent mass in the equations of motion of this $2+1$ dimensional model deserves comment.
At $t=0$ the effective mass is positive and the system fluctuates around the false vacuum $|\vec{\Phi}|=0$. When
t reaches $t_c$ the system is destabilized and the fields begin to roll to the true vacuum $|\vec{\Phi}|=1$. 
This behavior mimics that of a phase transition in $3+1$ spacetime dimensions.  It is well-known 
that there is no continuous symmetry breaking and accompanying long range order in spatial dimensions
of two or less \cite{MerWag}.  Therefore the time dependence of the effective mass does {\it not} come
from changing a thermodynamic variable, such as temperature.  However, though the $2+1$ dimensional action of 
Eq. (\ref{eq-action}) has no finite-temperature equilibrium phase transition there can still be an {\it effective} dynamical
transition.  When energy is dissipated, a high temperature initial state dominated by short wavelength thermal 
fluctuations will eventually evolve into a state dominated by long wavelength topological textures.  Textures become 
quasi-stable excitations when the effective temperature drops substantially below
the texture energy of $8\pi$.  The use of an external time-dependent effective mass allows for
a controlled passage between the fluctuation-dominated and texture-dominated phases.  Within this controlled
setting the role of the quench rate in the formation of texture can be isolated and studied effectively.  The lessons
learned are applicable to more realistic phase transitions, both in the early universe and in the laboratory.

\subsection{Numerical parameters and techniques}
The dynamics of the non-linear stochastic system was solved numerically.  Eq. (\ref{eq-fieldeq2}) was discretized on
a 2-dimensional lattice using a standard second-order leapfrog method for the time evolution 
\cite{Tur2,AntBet,PresRydSpergel} 
and a second-order spatial discretization for the laplacian $\bigtriangledown^2$,

\begin{eqnarray}
\vec{\Pi}_{n+1}[j][k]& = &\frac {1-\chi}{1+\chi}\vec{\Pi}_n[j][k]+ \frac{dt}{(1+\chi)}\nonumber \\
& &  (\frac{1}{dx^2}(\vec{\Phi}_n[j+1][k] -4\vec{\Phi}_n[j][k]\nonumber \\
& & +\vec{\Phi}_n[j-1][k]+\vec{\Phi}_n[j][k+1] \nonumber \\
& &  +\vec{\Phi}_n[j][k-1])-\vec{\Phi}_n[j][k](\vec{\Phi}_n^2[j][k]   \nonumber \\
& &  +m_{eff}^2[ndt])+\xi_n[j][k]) \nonumber
\end{eqnarray}
\begin{eqnarray}
\vec{\Phi}_{n+1}[j][k]=\vec{\Phi}_n[j][k]+dt\vec{\Pi}_{n+1}[j][k] \\
& & \nonumber 
\end{eqnarray}

\noindent where $\vec{\Pi}=\frac{\partial\vec{\Phi}}{\partial t}$ and $\chi=\eta dt$.  The labels $[j][k]$ identify 
the lattice point and $n$ labels the time step. At each time step and each lattice point the noise 
is generated through a sum of $M$ randomly distributed numbers \cite{AlexHabKov}
\begin{equation}
   \xi[j][k]=\Sigma_{i=1}^M{\frac{\theta[j][k]}{M} (\frac {24M\eta T} {dx^2dt})^{\frac{1}{2}}},
\end{equation}      
\noindent where $-0.5<\theta[j][k]<0.5$ are a set of $N^2$ random numbers.  In the limit when
$M\rightarrow\infty$ the $\xi[j][k]$ distribution approaches that of a Gaussian with the variance required for
the fluctuation-dissipation relation Eq. (\ref{eq-fd}).  In practice $M=24$ was used and no change in the dynamics was
observed for larger $M$.  The timestep was $dt=0.02$ and the lattice spacing was fixed at $dx=0.5$ on a 
square lattice with $N=1000$ sites per side.
There are three physical parameters in the system, the quench rate $\tau$, the dissipation constant $\eta$,  
and the initial temperature $T$.  The quench rate was varied between $\tau=10$ and $\tau=150$.  
Faster quenches (smaller $\tau$) deviate from the observed power-law behavior and asymptote 
to the behavior characteristic of an instantaneous quench as has been previously observed \cite{ZurYat}.  Longer quenches
are in principal possible but take substantially more computer time.  The dissipation constant was chosen at
$\eta=1.0$ to produce relatively overdamped behavior.  The evolution equations were supplemented with initial conditions
generated by solving the equilibrium 
dynamics of Eq. (\ref{eq-fieldeq2}) with a value of
the effective mass {\it fixed} at one quench timescale from the critical point,
\begin{equation}
 m_{eff}^2(t<0)= \tanh{(1.0)}.
\end{equation}
\noindent Run times for the system to reach an order parameter of $\phi=0.95$ on a DEC 500 Mhz workstation varied 
from hours for $\tau=10$ to days for $\tau=150$.

\subsection{Properties of the texture distribution}

Field configurations obtained from numerical simulations provide an embarrassment of riches.  In order to explore fully the 
participation of topological textures in the dynamics of the phase transition it is necessary to cull from the 
abundance of information contained in the snapshots of the topological field configurations, a few 
well-chosen functions whose time evolution provides a clear 
indication of the dynamics.  The late-time coarsening of a textured system is described by at least 
{\it three} different length scales \cite{ZapZak,Rut}, the average defect-defect separation $L_{sep}$, the average 
defect width $L_w$, and the average texture-antitexture separation $L_{tat}$.  In the formation dynamics studied here
we were able to measure reliably only $L_{sep}$ and $L_w$.  These length scales are defined through the 
the topological charge density and the topological defect two-point correlation function.   
Consider the two-point defect density correlation function

\begin{equation}
C_1(r,t)=\langle |\rho(\vec{x})||\rho(\vec{x}+\vec{r})| \rangle -\langle|\rho(\vec{x})|\rangle ^2,
\end{equation}

\noindent where $\langle \rangle$ denotes average over the lattice.  Define the normalized correlation function

\begin{equation}
G(r,t)\equiv \frac {C_1(r,t)} {C_1(0,t)}.
\end{equation}

\noindent $G(r,t)$ is a statistical measure of the ``lumps'' in the defect density function $|\rho(\vec{x})|$. 
The average texture width $L_w$ is proportional to the half-height scale $r_{\frac{1}{2}}$ of $G(r,t)$ defined by

\begin{equation}
G(r_{\frac{1}{2}}, t)=1/2.
\end{equation}

\noindent To measure the average texture separation consider

\begin{equation}
Q_{total}=\int d^2x |\rho|.
\end{equation}

\noindent $Q_{total}$ counts the total number of topological defects in the system without distinguishing 
between textures and antitextures.  On dimensional grounds,

\begin{equation}
Q_{total}\sim \frac{1}{L^2_{sep}}L_{sys}^2,
\end {equation}

\noindent where $L_{sep}$  measures the average separation of the topological defects and $L_{sys}$ is
the physical size of the lattice which did not vary between runs.   These two length scales, $L_{w}$and $L_{sep}$, 
provide a detailed characterization of the topological texture network at the end of the transition. In principle, each scale 
offers a different window into the nonequilibrium phase transition.

\subsection{The dynamics of symmetry breaking}
\begin{figure}[htb]
\epsfig{file=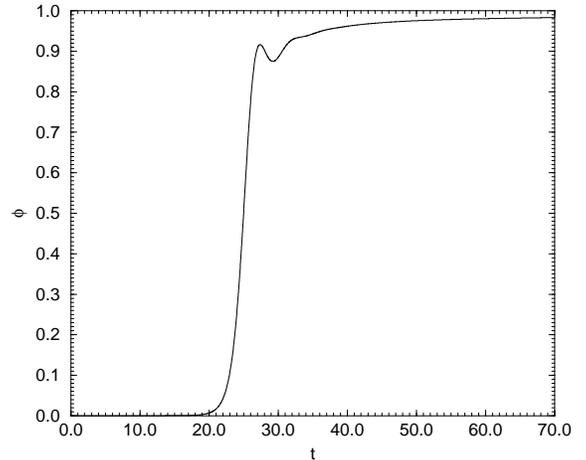, width=3.4 in}
\caption{Plot of the order parameter $\phi(t)=\langle |\vec{\Phi}|^2 \rangle$ 
versus time $t$ for quench parameter $\tau_q=10.0$.  Although the system is overdamped
near freeze-out, the small oscillations in $\phi$ near the end of the transition indicate that 
oscillations around the true vacuum are not completely overdamped.} 
\label{fig-order}
\end{figure}

\begin{figure}[htb]
\epsfig{file=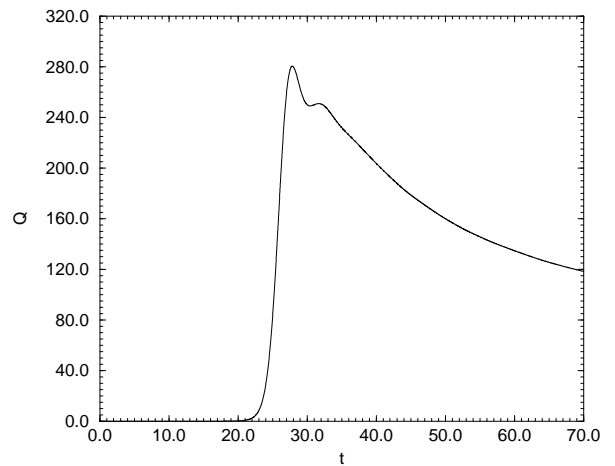, width=3.4 in}
\caption{Plot of the number of textures  $Q(t)=\int d^2x |\rho(\vec{x},t)|$ 
 versus time $t$ for quench parameter $\tau=10.0$.  As with the order parameter, 
the small ''bump'' indicates oscillations around the true vacuum near the end of the transition} 
\label{fig-charge}
\end{figure}

Plots of the time evolution of the order parameter $\phi\equiv \langle\vec{\Phi}^2 \rangle$
and the total number of topological defects $Q_{total}=\int d^2x \mid \rho(x,t) \mid$ are shown in figures 
(\ref{fig-order}) and (\ref{fig-charge}) respectively.  The plots correspond to an evolution with
quench parameter $\tau_q=10$.  In the symmetric state
$(t<10.0)$ both $\phi$ and $Q_{total}$ are close to zero.  When $t>10.0$ the external effective mass is negative
and the fields, kicked by the small stochastic force $\xi(\vec{x},t)$, begin to roll to the true vacuum $\phi=1$.
The order parameter $\phi$ is a measure of the number of lattice points where the fields have fallen to the true vacuum. 
As $\phi$ increases both the gradient energy and the number of defects increase.  However as $\phi\rightarrow 1$ most of
the lattice is near the ground state vacuum and the defect formation process is over.  After formation, the defect 
density does not remain constant but decreases slowly as texture-texture and texture-antitexture interactions dominate.

\section{Unraveling critical dynamics}
\begin{figure}[htb]
\epsfig{file=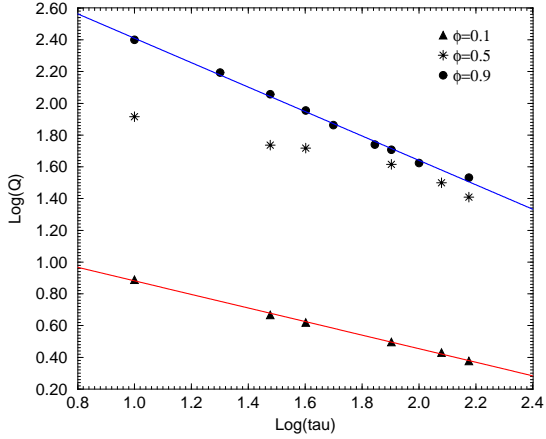, width=3.2 in}
\caption{Plot of the average number of textures determined from the topological charge  
$Q=\int d^2x |\rho(\vec{x},t)|$ versus quench parameter $\tau$ for
different values of the order parameter $\phi=\langle \vec{\Phi}^2 \rangle$.  Symbols denote lattice
measurements and the two solid lines denote best-fit power laws.  Early in the phase transition $\phi=0.1$
the best-fit power law is $\log(Q)=(1.31\pm 0.02)- (0.43 \pm 0.01 )\log(\tau)$, in good agreement with the 
Kibble-Zurek mechanism.
At late times $\phi=0.9$, the best-fit power law is, $\log(Q)=(3.17 \pm 0.03) - (0.77\pm 0.02)\log(\tau)$.  
Without texture interactions, the early and late-time power laws would be the same.  However in this
interacting system, texture unwinding and texture-antitexture annihilation steepen the power law observed
at the end of the phase transition.  The crossover is apparent at intermediate times $\phi=0.5$ as defect
interactions modify the number of textures for longer quenches.}
\label{fig-Qevolution}
\end{figure}

\begin{figure}[htb]
\epsfig{file=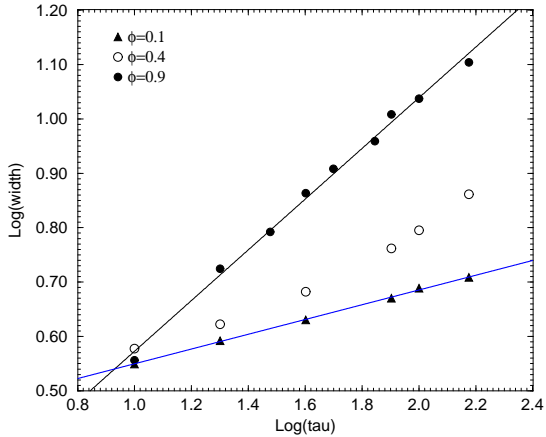, width=3.2 in}
\caption{Plot of the average texture width versus quench parameter $\tau$ for different values of the
order parameter $\phi=\langle \vec{\Phi}^2 \rangle$.  
Symbols denote lattice measurements and the two solid lines denote best-fit power laws.  Early in the phase transition $\phi=0.1$
the best fit power law is $\log(L_w)=(0.42\pm)(0.14\pm 0.01)\log(\tau)$.
At late times $\phi=0.9$, $\log(L_w)=(0.11\pm 0.02) - (0.47\pm 0.01)\log(\tau)$.  
At intermediate times $\phi=0.4$, $L_w$ is growing at a faster
rate for the longer quenches.  The early-time power law exponent is smaller than expected from the Kibble-Zurek mechanism.
However, at early times the form of defect-defect correlation function.}
\label{fig-Qwidthevolution}
\end{figure}

Figures (\ref{fig-Qevolution}) and (\ref{fig-Qwidthevolution}) show Log-Log plots of the average number of textures and 
the average texture width versus the quench rate.  The top graph in each figure displays the average number of textures and the
average texture width measured near the end of the transition, taken to be the point where
$\langle \vec{\Phi}^2 \rangle =0.9$.  The data are well-described by the power-law scalings 
\begin{eqnarray}
\label{eq-latescale}
L_{sep} \sim \tau^{0.38 \pm 0.01} \\
L_{w}    \sim \tau^{0.47 \pm 0.01} \nonumber
\end{eqnarray} The best-fit exponents were derived using equal weighting for all lattice measurements.
As we found no reliable means to estimate the error in individual lattice measurements, 
the errors in the scaling exponents arise from statistical errors in the fit.  
Certainly, power-law scaling is expected from the freeze-out picture.  However, the Kibble-Zurek mechanism 
also provides a precise prediction for the {\it value} of the power-law exponent as derived from equilibrium 
critical exponents.  The model under consideration is a {\it mean field} model and the only role of the
small fluctuations is to seed texture formation.  Therefore it is expected that $\nu=\frac{1}{2}$,
the mean field value of equilibrium correlation length exponent.  This expectation was verified
by equilibrium lattice simulations.  With $\eta=1.0$ and $\tau \geq 10$ the dynamics is overdamped
at freeze-out.  Therefore $\mu=2\nu=1$ and the Kibble-Zurek mechanism predicts the scaling
$\xi(\tau) \sim \tau^{\frac{1}{4}}$, in obvious disagreement with Eq. (\ref{eq-latescale}).         

The larger observed exponent means that for longer quenches topological textures are both less numerous and bigger
than the Kibble-Zurek mechanism predicts.  However, slower quenches take longer for the phase transition to complete  
since as $\tau$ increases, the order parameter rolls more slowly. In fact, observations of the dynamics show that the 
time $t_{0.9}$ (measured from the critical point) it takes for the order parameter to reach 
$\langle \vec{\Phi}^2 \rangle =0.9$ increases with $\tau$ in a simple way $t_{0.9} \sim 1.5 \tau$.
With more time to evolve between freeze-out and the end of the transition, slower quenches allow   
for more interactions among the texture distribution.  This suggests that the apparent power-laws observed at
the end of the transition are {\it not} indicative of a single scale but a combination of the Kibble-Zurek 
mechanism and the dynamical length scales of the evolving texture network.

\begin{figure}[htb]
\epsfig{file=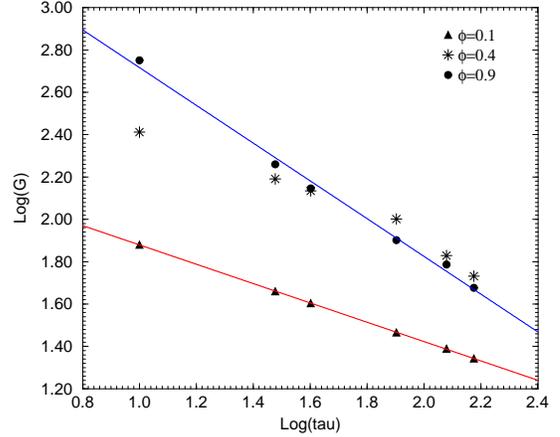, width=3.2 in}
\caption{Plot of the average number of textures determined from the gradient energy
$G=\frac{1}{16\pi}\int d^2x (\vec{\partial}\Phi_i \cdot \vec{\partial}\Phi_i)$
versus quench parameter $\tau$ for different values of the order parameter $\phi=\langle \vec{\Phi}^2 \rangle$.  
Symbols denote lattice measurements and the two solid lines denote best-fit power laws.  Early in the phase transition $\phi=0.1$
the best fit power law is $\log(G)=(2.34 \pm 0.01)- (0.46 \pm 0.01)\log(\tau)$, in agreement with the Kibble-Zurek mechanism.
At late times $\phi=0.9$, $\log(G)=(3.65\pm 0.10) - (0.94\pm 0.06)\log(\tau)$. As in figure \ref{fig-Qevolution} a crossover is
apparent at intermediate times $\phi=0.4$.  The late-time power law is slightly steeper than that determined from the
topological charge because the gradient energy also includes topologically trivial configurations which
decay more rapidly than textures.}
\label{fig-Gevolution}
\end{figure}

Evolution of the power law exponents of the texture length scales is clearly indicated in the two lower graphs of
figures \ref{fig-Qevolution} and \ref{fig-Qwidthevolution}.  The lowest graph in each figure 
displays the average number of textures and the average texture width measured very early in the
transition when  $\langle \vec{\Phi}^2 \rangle =0.1$.  At these early times the effect of coarsening
dynamics is insignificant and power-law scaling indicative of the Kibble-Zurek mechanism is evident,
\begin{eqnarray}
\label{eq-earlyscale}
L_{sep} \sim \tau^{0.22 \pm 0.01}, \\
L_{w}    \sim \tau^{0.14 \pm 0.01}. \nonumber
\end{eqnarray}
\noindent Although the power law exponent of $L_w$ is slightly smaller than predicted by the Kibble-Zurek
 mechanism, the difference is not likely to be significant.  Textures are not well-formed at these early times 
and the method for measuring $L_w$ is sensitive to the details of the two-point function.    
In the middle graphs of each figure, the number of textures
and the texture width were measured at intermediate times.  Both show the crossover from
formation dynamics to coarsening dynamics as longer quenches develop a steeper power-law indicative
of texture interactions.  A similar picture of the evolution of $L_{sep}$ is given by the total
gradient energy \ref{fig-Gevolution}.

\begin{figure}[htb]
\epsfig{file=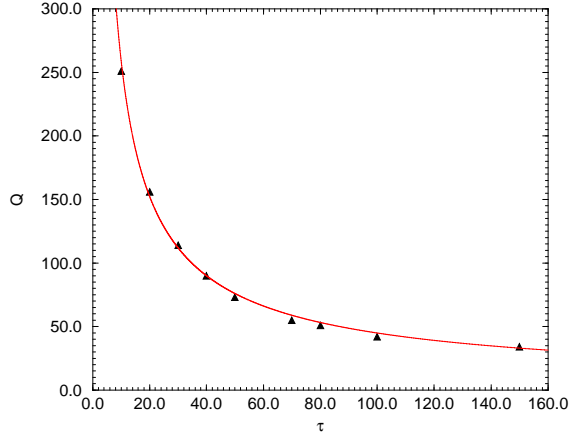, width=3.2 in}
\caption{Plot of the average number of textures $Q=\int d^2x |\rho(\vec{x})$ versus
quench parameter $\tau$.  For each quench the number of textures was measured 
at the point when $\langle \vec{\Phi}^2 \rangle=0.9$.  Filled triangles denote lattice
measurements.  The solid line denotes the theoretically expected number of textures computed
from the length scale $L_{sep}(t)=\xi_{freeze}+ (\xi_{freeze})^{\frac {1}{3}}(t-t_{freeze})^{\frac {1}{3}}$.
The freeze-out correlation length determined from this fit is $\xi_{freeze} \sim \tau^{0.24 \pm
0.02}$, in excellent agreement with the Kibble-Zurek mechanism. } 
\label{fig-Lsepfit}
\end{figure}

\begin{figure}[htb]
\epsfig{file=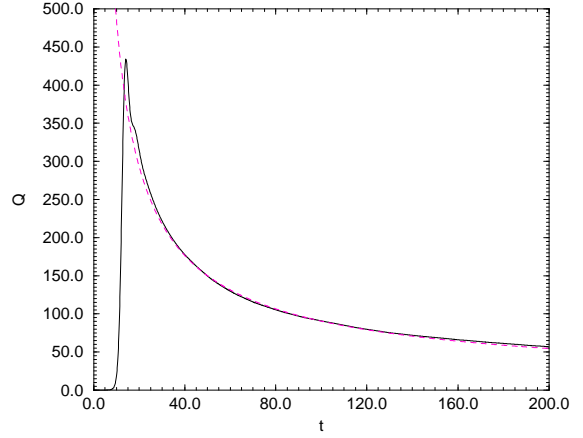, width=3.2 in}
\caption{Plot of the average number of textures $Q(t)=\int d^2x |\rho(\vec{x},t)|$ versus time for a very fast quench $\tau=1$.
The solid line denotes data from a numerical simulation.  The dashed line is the best-fit power law of the late-time coarsening
dynamics, $Q(t) \sim t^{-0.72}$.  The coarsening exponent $\alpha=-0.72$ agrees favorably with previous studies which found
$\alpha=-\frac{2}{3}$}
\label{fig-longcharge}
\end{figure}

Figures \ref{fig-Qevolution} and \ref{fig-Qwidthevolution} provide clear evidence of the evolution of
the power-laws characterizing the length scales of the texture distribution.  It is also possible to use 
the knowledge of
texture coarsening dynamics to provide a {\it quantitative} understanding of the power-laws observed
at the end of the transition.  As discussed in section IIA, the phase ordering of a textured system results in a 
dynamical length scale characterizing the separation of defects, $L_1(t)=\xi_0^\frac{1}{3}t^\frac {1}{3}$.  The
dynamics of this length scale for a fast quench (which enters the coarsening regime earlier) is shown in
Figure \ref{fig-longcharge}.  In the instantaneous quenches of \cite{Rut} $\xi_0$ was identified 
with the initial correlation length.  In a dynamical
quench it is more natural to identify $\xi_0$ with the freeze-out correlation length.  Combining the Kibble-Zurek
mechanism with phase ordering dynamics I suggest that the evolution of the scale characterizing the separation of 
topological textures is given by

\begin{equation}
\label{eq-Lsep}
L_{sep}(t)=\xi_{freeze}+ (\xi_{freeze})^{\frac {1}{3}}(t-t_{freeze})^{\frac {1}{3}},
\end{equation}

\noindent where $\xi_{freeze} \sim \tau^\alpha$ is the length scale determined by the Kibble-Zurek
mechanism, $t_{freeze} \sim \tau^{2\alpha}$ is the time from ``freeze-out'' 
and t is the time from the critical point. Although Eq. (\ref{eq-Lsep}) is slightly complicated, its
form is tightly constrained and the only free parameter is $\alpha$, the exponent of the freeze-out
correlation length.  When $t=t_{0.9}\sim 1.5\tau$, the expected number of defects 
$Q \sim \frac{1}{L^2_{sep}}$ is matched against the number of defects determined from the 
numerical simulations.  A plot of the data and the best-fit length scale is shown in figure \ref{fig-Lsepfit}.  
To obtain the fit one data point is used to normalize the scale and 
a value of $\alpha$ is determined by minimizing $\chi^2$.  By using, in turn, each data point as a normalization
a spread in the value of $\alpha$ is obtained.  The best-fit value of $\alpha$ is taken as the average over all
normalizations and the error is the standard deviation.  As determined from these fits, $\alpha=0.24 \pm 0.02$ 
in excellent agreement with $\alpha=0.25$ expected from the Kibble-Zurek mechanism.  

\begin{figure}[htb]
\epsfig{file=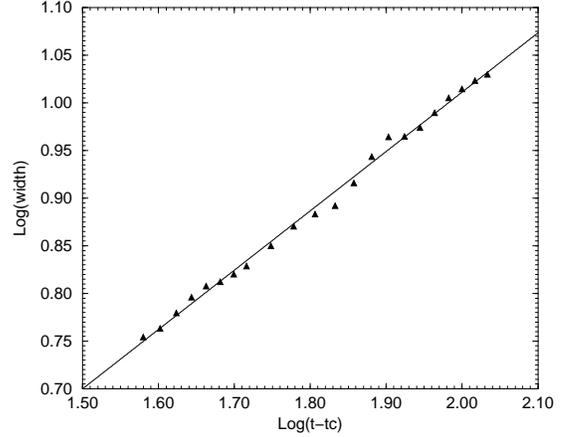, width=3.2 in}
\caption{Log-Log plot of the texture width versus time for quench parameter $\tau=80$.  Time is measured from
the critical point which occurs when $t_c=80$.  Solid triangles denote data from a numerical simulation.  The straight  
line is the best-fit power law of the coarsening dynamics, $L_w(t-t_c) \sim (t-t_c)^{-0.62}$.  The coarsening exponent
$\alpha=-0.62$ is close to the simple scaling exponent $\alpha=-\frac{1}{2}$.}
\label{fig-longlogwidth}
\end{figure}

The dynamics of $L_w$, the length scale characterizing the size of topological textures, is more
complicated.  Figure \ref{fig-longlogwidth} provides an example of texture width dynamics for $\tau=80$. 
In the simulations the texture width was observed to grow in time with a power law
near the simple coarsening dynamics predicted for an overdamped model $L \sim t^{\frac{1}{2}}$. 
However, this growth rate is different from previously reported studies of the late-time dynamics 
of the texture width.  Consistent with previous studies
\cite{Rut} it is possible that $L_w(t)$ reaches its asymptotic dynamics only at later times.  It is also possible 
that $L_w(t)$ is contaminated at early times by topologically trivial spin configurations whose decay then dominates 
the dynamics.  The
resolution of this issue requires a more accurate method of counting textures e.g. by looking explicitly at
the winding around the vacuum manifold.  Whatever the mechanism, Figure \ref{fig-longlogwidth} shows 
that $L_w$ can evolve significantly between the early and late periods of the phase transition.  
As a concrete equation,  I suggest that the evolution is given by

\begin{equation}
\label{eq-Lwidth}
L_{w}(t)=\xi_{freeze}+ (t-t_{freeze})^{\frac{1}{2}}.
\end{equation} 

\begin{figure}[htb]
\epsfig{file=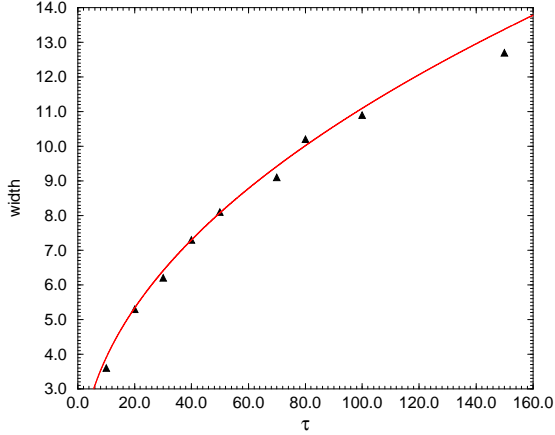, width=3.2 in}
\caption{Plot of the average texture width (in lattice units) versus
quench parameter $\tau$. For each quench the average width was measured 
at the point when $\langle \vec{\Phi}^2 \rangle=0.9$.  Filled triangles denote lattice
measurements.  The solid line denotes the theoretically expected width computed
from the length scale $L_{w}(t)=\xi_{freeze}+ (t-t_{freeze})^{\frac{1}{2}}$.
The freeze-out correlation length determined from this fit is $\xi_{freeze} \sim \tau^{0.22 \pm
0.07}$, in agreement with the Kibble-Zurek mechanism. } 
\label{fig-Lwidthfit}
\end{figure}

\noindent $\xi_{freeze} \sim \tau^\beta$ is the length scale determined by the Kibble-Zurek
mechanism, $t_{freeze} \sim \tau^{2\beta}$ is the time from ``freeze-out'' 
and t is the time from the critical point.  As before, when $t=1.5\tau$, the expected width of the textures is
matched against the width determined from the numerical simulations.  A plot of the data 
and the best-fit length scale  is shown in figure \ref{fig-Lwidthfit}.  The fits were obtained in the 
same way as for $L_{sep}$.  As determined from these fits, $\beta=0.22 \pm 0.07$ 
also in agreement with $\beta=0.25$ expected from the Kibble-Zurek mechanism.  

It is clear from Figs. (\ref{fig-Lsepfit}) and (\ref{fig-Lwidthfit}) that the power-laws
of equation (\ref{eq-latescale}) can be explained by incorporating texture dynamics.  Eqns. \ref{eq-Lwidth}
and \ref{eq-Lsep} offer a conservative, simple guess to the proper combination of formation and coarsening dynamics.  
A more complete analysis, with better techniques for identifying textures, would fit $L_{sep}(t)$ and $L_w(t)$ 
to functions with both general formation {\it and} coarsening exponents.  Nonetheless it is remarkable that these
simple equations are in good agreement with the Kibble-Zurek mechanism. Even if the exact form of 
Eqns. (\ref{eq-Lsep}) and (\ref{eq-Lwidth}) changes,  the observation remains that the length scales characterizing
the topological defect distribution at the end of the phase transition are a combination of defect interaction
and formation dynamics. 

It is not surprising 
that defect interactions can influence the properties of the defect network
before the end of the transition.  Although the long-wavelength dynamics is very slow until after the 
freeze-out time $t_{freeze} \sim {\tau}^{\frac{1}{2}}$ (in the overdamped
case), the time to complete the phase transition is $t_{0.9} \sim \tau$ allowing plenty of time for 
even partially formed defects to influence each other.  In previous numerical simulations in 1 and 2 
dimensions \cite{ZurLag,Jac,ZurYat} this effect was not noticeable (in overdamped evolutions)
because the interactions between defects were extremely weak. 
In a recent simulation of global vortex formation in 3 dimensions \cite{AntBetZur} one might expect an observable change in 
the scaling exponents since the defect interactions are much stronger.  However,
the transition was at relatively high temperature and the defects were at least partially 
screened by thermal fluctuations.  

\section{Texture Coarsening and the Kibble-Zurek mechanism}

The close agreement between Eqns. (\ref{eq-Lsep}) and (\ref{eq-Lwidth}) and the length scales observed at the end of the 
transition provides evidence that the {\it a priori} independent length scales, $L_{sep}$ and $L_{w}$, are formed
by the same nonequilibrium dynamics of the Kibble-Zurek mechanism.  That this should be so is not 
obvious.  In fact, the Kibble-Zurek mechanism can be questioned even in models which at late times depend 
only upon a single scale.  In \cite{KarRiv} it was argued
argued that the length scales defined by the field correlation length and vortex 
density, in principle, derive from different attributes of the power spectrum.  
Only under restricted circumstances does the defect separation follow the field correlation length
and scale with the quench rate.  However, the quench considered here was
deep into the spinodal region and the early-time power spectrum is dominated by
a single peak with momentum $k=k_{max}$ in the low-momentum modes.  When the power spectrum
is strongly peaked, the freeze-out correlation length $\xi_{freeze} \sim \frac{1} {k_{max}}$
determines a unique scale at early times to which all other length scales are connected.    

\subsection{Late-time dynamics and Experiments}
The strong violation of the scaling hypothesis observed in the coarsening of $2+1$ dimensional
textured systems suggests the possibility of novel, late-time measurements of early-time
critical dynamics.  As discussed in section (IIA), the coarsening of a textured system
with an instantaneous quench retains ``memory'' of a length scale $\xi_0$, the correlation
length in the disordered phase.  In a nonequilibrium phase transition it is the freeze-out 
correlation length, determined by the Kibble-Zurek mechanism, and {\it not} the correlation 
length of the initial state, that scars the texture distribution at late times.  For overdamped coarsening 
dynamics at late times,

\begin{equation}
\label{eq-latetime}
\frac{\xi^2_w(t)}{\xi_{sep}(t)} = constant=\xi_{freeze}.
\end{equation}  

\noindent The freeze-out correlation length can be measured by a late-time measurement of the (average) 
texture width and texture separation.  It is the particular (and not well-understood) dynamics of 
texture-texture interactions that produces strong scaling violations and unusual memory effects.  
In contrast, experimental probes of vortex formation in nonequilibrium phase transitions of $^3He$ and $^4He$ 
are hampered by the details of the interacting vortex distribution.  Because these systems approach a scaling solution
at late times, vortex interactions work to erase any memory of the initial state.  This is a 
particular problem in the $^4He$ experiments where the vortex network is experimentally observable only at later times and 
the number of initial defects must be inferred using detailed assumptions about the decay of the vortex tangle.

Textured systems in $2+1$ dimensions can be created in the laboratory.  If their dynamics are
approximately described by Eq. (\ref{eq-fieldeq2}) then late-time experimental measurements of
the Kibble-Zurek mechanism are possible.  Topological textures can also be formed during phase
transitions in $3+1$ dimensions in superfluid $^3He$ and the early universe.  In $3+1$
dimensions without additional higher derivative terms in the action, topological textures are unstable.
In this case the dynamics of coarsening is complicated by texture collapse.  
However, as long as textures exist in the system scaling violations and associated dynamical memory
are possible.    

\section{Conclusion}

In this work, we exploited the multiple length scales of a textured system to study a
nonequilibrium phase transition of an overdamped classical 0(3) model in $2+1$ dimensions.
At very early times, we identified power-law scaling characteristic of the Kibble-Zurek mechanism
and a single freeze-out scale.  This  
suggests that the multiple length scales characteristic of the late-time ordering of a 
textured system derive from the critical dynamics of a {\it single} nonequilibrium correlation length. 

When observed near the end of the phase transition, we found the scaling of the texture separation $L_{sep}(\tau)$ and 
the texture width $L_w(\tau)$ results instead from a competition between the length scale determined at freeze-out 
and the ordering dynamics of a textured system. We expect that this
observation is not restricted to systems of 
topological textures or to low dimensions but is relevant anytime defect
interactions are significant.   It is not surprising that defect interactions can significantly
modify the defect distribution even before the end of the phase transition. Well before defects are fully formed 
they frustrate the system from its true minimum energy ground state and interact with eachother.

The Kibble-Zurek mechanism provides a useful connection between critical dynamics 
and the length scales of the topological defect distribution observed at the end of the phase transition.
However, power-law scaling provides a rough and rather opaque window into
the complicated nonequilibrium processes of the phase transition. As demonstrated explicitly here, power-law scaling 
can also arise from completely different mechanisms such as defect interactions.  To understand the characteristics
of the topological defect distribution at the end of the transition it is therefore important to look closely 
both at the nonequilibrium dynamics of the phase transition and the evolution of the defect network.

\section*{Acknowledgments}
This work is supported in part by the National Science Foundation under grant PHYS-9800967.  
I would like to thank Esteban Calzetta and Bei-Lok Hu for very useful criticisms and interesting discussions.
I would also like to thank the National Scalable Computer Project at the University of Maryland for access to computing
resources.

%%%%%%%%%%%%%%%%%%%%%%%%%%%%%%%%%%%%%%%%%%%%%%%%%%%%%%%%%%%%%%%%%%%%%%%%%%%%
%                                                                          %
% Figures                                                                  %
%                                                                          %
%%%%%%%%%%%%%%%%%%%%%%%%%%%%%%%%%%%%%%%%%%%%%%%%%%%%%%%%%%%%%%%%%%%%%%%%%%%%
\newpage

%\begin{figure}[htb]
%\epsfig{file=lattice.eps, width=3.2	 in}
%\caption{Lattice snaphot of the toplogical charge density 
%$\rho(\vec{x},t)=\frac{1}{4\pi}\vec{\Phi}\cdot (\partial_x\vec{\Phi}\wedge \partial_y\vec{\Phi})$.
%This representative snapshot was taken for an evolution with $\tau_q=20$ at the point when $\langle \vec{\Phi}^2 \rangle =0.9$.
%Horizontal axes label the location in the lattice while the vertical axis labels
%the charge density $\rho(\vec{x},t)$ (in lattice units).  Only a small number of the $1000^2$ lattice points are shown.
%At this point in the evolution the sytem contains a large number of positive winding textures
%(peaks in $\rho(\vec{x},t)$) and negative winding textures (valleys in $\rho(\vec{x},t)$).}
%\label{fig-lattice}
%\end{figure}

\newpage

\end{document}